\documentclass[apj,twocolumn]{openjournal}
\usepackage{amsmath}
\usepackage{orcidlink}
\usepackage{booktabs}
\usepackage{multirow}
\usepackage{color}
\usepackage{soul}
\usepackage{hyperref}
\hypersetup{
	colorlinks=true,
	urlcolor=blue,    
	linkcolor=blue,  
	citecolor=blue,   
}

\newcommand{\orcidauthor}[3]{\author{\href{http://orcid.org/#1}{#2$^{#3}$}}}

\begin{document}


\title{On Benchmarking SRc $\alpha$ Ori using Period-Luminosity Relationship} 




\shorttitle{CGHA and WADS}
\shortauthors{CGHA and WADS}

\thanks{$^*$Corresponding E-mail: \href{mailto:genhcapuli@rtu.edu.ph}{genhcapuli@rtu.edu.ph}}

\orcidauthor{0000-0003-1253-7043}{Generich H. Capuli}{*}

\author{Willie Anthony D. Sapalaran}
\affiliation{Department of Earth and Space Science (DESS), Rizal Technological University, Boni Ave, Brgy. Malamig, Mandaluyong City, \\ Metro Manila, 1550}




\begin{abstract}

We conducted a benchmarking analysis of the semi-regular pulsator and red supergiant $\alpha$ Ori. In its dimming episode last 2020, our observational results include the binned measurements from the space-based telescope SMEI collated. We report a long secondary period of $P_{\text{LSP}}$ = 2350 $\pm$ 10 d and a fundamental mode of pulsation of $P_0$ = 415 d $\pm$ 30 d, with a radial velocity amplitude of the FM at $A_{RV}$ = $2.21_{-0.50}^{+0.95}$ km s$^{-1}$. Meanwhile, we also detected the first overtone component of $P_1$ = 185 d. The derived harmonics of $\alpha$ Ori, including this newly acquired variation, support and align with the current literature.

At $\sim$2.20 $\pm$ 0.10 $\mu$m, we acquired Near-Infrared $K$-band photometric measurements from several catalogs and surveys, calibrated accordingly. Our assigned inherent color lies midway between the extremes reported in existing literature. Additionally, we determined a weighted excess color index of $E_{(B-V)}$ = 0.340, and using a $K$-extinction factor of $R_K$ = 0.382, we derived an extinction of $A_K$ = 0.130. By subtracting extinction from all $K$-band photometric measurements and applying both the linearity and the newly derived distance from previous studies, our analysis yields a luminosity of log($L$/$L_{\odot}$) = 5.00 $\pm$ $0.15_{(-0.45)}^{(+0.48)}$ for $\alpha$ Ori. 

In turn, this allowed us to conduct the benchmarking scheme alongside the data from existing reports that are stitched together using Period-Luminosity ($P$-$L$) relationship. This results in a best-fit relation of log($L$/$L_{\odot}$) = (7.26 $\pm$ 0.10) log $P$ + ($-14.10$ $\pm$ 0.25) and reveals that $\alpha$ Ori can be situated at the lower bound of the 18 $M_{\odot}$ regime due to current pulsation trends.

\keywords{stars: semi-regular variables --- stars: individual ($\alpha$ Ori) --- 
stars: pulsation --- stars: luminosity --- stars: period-luminosity}

\end{abstract}

\maketitle 
\section{Introduction}\label{intro}

Classically, Red Supergiants (RSGs) are Population I massive (10 $\geqslant$ \textit{M}\textsubscript{init} $\leqslant$ 30 $M_{\odot}$) descendants of OB-type main-sequence (MS) stars in the core Helium burning (CHeB) period, preceding as core-collapse Type II supernovae (CCSNe); either II-P(lateau) or II-L(ong decline), leaving behind as dense neutron stars (NS). Though, some RSGs may undergo a 'blue loop' before its final death plunge \citep{DaviesBeasor.2018,Neugentetal.2020}. These late-type stars are allocated in the upper region of Hertzsprung-Russell Diagram (HRD) characterized with ages at ${\sim}$ 8 -- 20 Myr, highly luminous i.e. 10$^{5}$ -- 10$^{6}$ $L$/$L_{\odot}$ at the expense of having cool effective temperature (\textit{T}\textsubscript{eff}) ranging from 3000 -- 4500 K, giant stellar radii typically at 100 -- 1500 $R_{\odot}$ and enormous mass loss rates (M\textsubscript{L}) of ${\sim}$ 10$^{-7}$ -- 10$^{-4}$ $M_{\odot}$/yr. With those properties and going through their lifetimes, these stars contribute significantly to the chemical enrichment of young stellar populations and for testing stellar evolutionary models supplemented by observational analysis \citep{Masseyetal.2007,MauronJosselin.2011,Meynetetal.2015,MasseyEvans.2016,Daviesetal.2017}.

Like other RSGs, $\alpha$ Ori ($\alpha$ Orionis, HD 39801) --- popularly known as Betelgeuse --- is no exception as it provides wealth of information to tackle on the intricate evolution of massive stars. Designated in the HRD and M-K spectral system as M2-Iab and a well-known Type \textit{C} semi-regular (SRc) variable, the bright pulsating RSG has been on the hot seat for popular press and astronomy aficionados caused by its deep dimming episode in November 2019 to March 2020 \citep{Guinan.2019,Dupreeetal.2020}. $\alpha$ Ori is characterized by a current mass of 16.5 -- 19 $M_{\odot}$ \citep{Joyceetal.2020}, \textit{T}\textsubscript{eff} of 3600 $\pm$ 25 K with a dilemma to the massive star's position on the HRD \citep{LevesqueMessey.2020}, varying size at ${\sim}$ 600 -- 1000 $R_{\odot}$, rectified proximity up to ${\sim}$ 130 -- 200 pc, and a promising candidate of being a post-merger star \citep{vanLeeuwen.2007,Harperetal.2008,Harperetal.2017,Dolanetal.2016,Joyceetal.2020,Chatzopoulosetal.2020}.

The term “\textit{semi-regular}” refers to the quasi-sequential fluctuations on its brightness mostly inferred as being caused by inconsistent cyclic variations. This points the finger at convection (including convective cells) to lie at the origin of inhomogeneities covering the simple pulsating photosphere of $\alpha$ Ori \citep{Chiavassaetal.2009,Chiavassaetal.2010,Chiavassaetal.2011}, which may upset a pressure ($\rho$-mode) oscillation in the low-overtone sequence driven in the Hydrogen (H) ionization region and cause the ${\sim}$ 300 -- 500 d of optical and UV flux modulation \citep{StothersLeung.1971,Goldberg.1984,Guin.1984,Dupreeetal.1987,Kissetal.2006,Chatysetal.2019,Joyceetal.2020}. Conversely, the brightness variation of $\alpha$ Ori that spans for about ${\sim}$ 2050 -- 2500 d likely falls under the subset of 'Long Secondary Periods' (LSP) and can be attributed to large cell turnover, rotational modulation, non-radial gravity mode (g-mode), magnetic activity, and binarity. Such behaviour has been observed on semi-regulars and RSG variables in the LMC. \citep{Kissetal.2006,Stothers.2010,SoszynskiUdalski.2014,PercyDeibert.2016}. In fact, \citet{Karovska.1986} proposed an unusual triple-star model for $\alpha$ Ori, a multiple stellar system, from their speckle-imaging measurements. Nearly 40 yrs after it was put forward, thorough and recent investigations both by \citet{Goldberg_2024} and \citet{MacLeod_2025} revealed that a binary low-mass companion, referred as $\alpha$ Ori B ($M_2$ $\leqslant$ 1.5 $M_{\odot}$), was the most plausible mechanism for the LSP behavior in $\alpha$ Ori. So far, this is consistent with observational analysis from Radial Velocity–Light Curve (RV-LC) phase offset. 

The variable star's proper motion within the medium is also captured by multiple space-based telescopes (e.g. AKARI IR, GALEX, Herschel, HST). In addition, these telescopes also revealed an asymmetric yet Oxygen (O)-rich bow shock arc with a linear bar around $\alpha$ Ori pointing in the direction of motion (at a position angle of ${\sim}$ $50^{\circ}$ E of N) and interpreted as the interface between the ISM and dusty CSE enriched by the M\textsubscript{L} ejections and stellar wind \citep{Mohamedetal.2012,Decinetal.2012,Mackeyetal.2013}. In fact, \citet{Coxetal.2012} found out that many AGBs and RSGs are shrouded by this structure ($\alpha$ Ori is Class I - Fermata). Still, this makes $\alpha$ Ori to appear spotty-patchy on the surface likely caused by stellar M\textsubscript{L} \citep{Hauboisetal.2009,Ohnakaetal.2009,Ohnakaetal.2011}. \citet{JosselinPlez.2007} asserts that convection, radiation pressure alongside velocity gradients, pulsation, shocks, and rotation are the leading processes that may play a role for triggering the M\textsubscript{L}.

In this paper, we used a variety of tools to delve into $\alpha$ Ori, observationally. We analyzed frequency or Fourier spectrum, and distinguish periods using photometric observations of the source. Secondly, we generate a Period-Luminosity (\textit{P}-\textit{L}) diagram for $\alpha$ Ori by calibrating the sourced Near-Infrared (NIR) \textit{K}-band magnitudes into Bolometric Luminosities (\textit{L}\textsubscript{bol}). Lastly, we conducted a benchmarking scheme with regards to $\alpha$ Ori's stellar \textit{M}, \textit{L}, and \textit{P}$_{\text{0}}$.  

This brief research report proceeds as follows: In Section 2, we discussed the point source’s light curve behavior through photometric observations from American Association of Variable Stars Observers (AAVSO), Solar Mass Ejection Imager (SMEI), and sourced throughout the NIR catalogues for \textit{K}-band. In Section 3, we presented the period analysis, \textit{L} calibration, and constructed a \textit{P}-\textit{L} proportionality for the interest. Lastly, we presented our results for $\alpha$ Ori along its best-fit, limitations, and possible future plans in Section 4.

\section{Observational Materials and Methods}

\subsection{Standard V-band Photometry}

\begin{figure*}[t!]
\centering
\includegraphics[width=\textwidth]{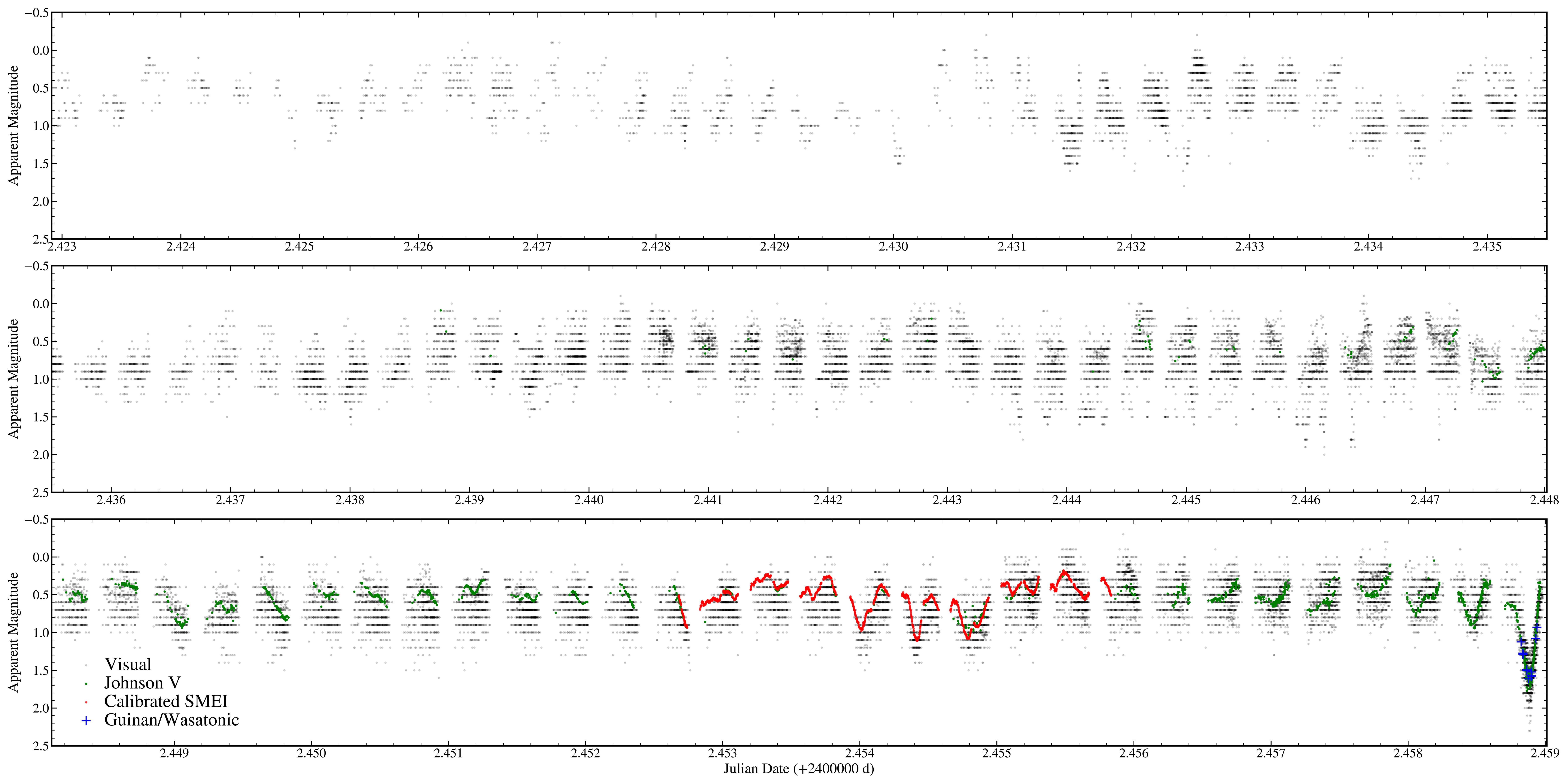}
\caption{Total light curve of $\alpha$ Ori combined from AAVSO's and SMEI's archive throughout the years. The GDE is standing tall as an outlier, while the rest varies at $\sim$1.0 mag. Also, included are the photometric observations conducted by Guinan and his collaborators \citep{Guinan.2019,Guinan.2020} during the GDE via The Astronomer Telegram. Each data are labeled accordingly as shown in the lower left.}
\label{fig:lightcurve}
\end{figure*}

Since our analysis includes only one pulsating RSG, we narrowed down our choice of archives to look through and use for this case. The long-term visual estimates of $\alpha$ Ori have varied over time both in quality and quantity. Brought by its stochastic and convective motions that drive pulsation, its circumstellar material, and M\textsubscript{L}, the semi-regularity of the star should produce noticeable photometric variations. Mostly, the American Association of Variable Stars Observers \citep[AAVSO;][]{Kafka.2020} have a collection of photometric observations by amateur and professional astronomers. They provide a well-defined scope for $\alpha$ Ori over the last 4 decades, including the recent deep dimming in mid-February 2020. We attempted to scour for additional resources in the All-Sky Automated Survey (ASAS) campaigns (ASAS-3 and ASAS-3N) at the widest aperture; MAG 4, to gather all the flux. However, the sensors are contaminated.

To supplement the AAVSO's photometric data, we utilized the Solar Mass Ejection Imager (SMEI) attached to the Coriolis spacecraft \citep{Jackson.2004}. Aside from detecting and forecasting heliospheric structures propagating from the Sun and a threat towards the Earth, the 104-minute cadence, centroid fitting method, and nullifying the bright sources in the raw heliospheric data can also be processed into a time-series photometric survey allowing SMEI to morph into one of the early space photometry missions \citep{Webbetal.2006,Buffingtonetal.2007,Hicketal.2007}. The raw photometric observations can be checked through their pipeline available by request to the head owners of SMEI, while \citet{Joyceetal.2020} provided a refined and single day-binned \textit{V}-band magnitudes extracted from the SMEI light curve for $\alpha$ Ori. Figure \ref{fig:lightcurve} depicts the total light curve of $\alpha$ Ori together with the AAVSO's and SMEI's photometric data from 1928 to mid-2020.

We used the Lomb-Scargle periodogram \citep{Lomb.1976,Scargle.1982} to calculate the power spectra and identify the periodic signals of unequally spaced time series. In order to analyze the full-scale time series, the Period04 software \citep{LenzBreger.2005} provides the tools needed to decompose the peak signal from the combined light curve and perform additional iteration in the residual data. The amplitudes are measured from the peak height within the Fourier spectrum giving the partial amplitude of the best-fitting sinusoidal wave. We note, however, that the confluence event (Great Dimming Event/GDE) was removed as it presents as an outlier to the time-series.

\subsection{Near-Infrared: K-band Photometry}

As previously mentioned in Section \ref{intro}, the analysis incorporates NIR \textit{K}-band magnitudes, with a wavelength of $\sim$2.20 $\pm$ 0.10 $\mu$m, to determine \textit{L}\textsubscript{bol} more accurately than using conventional \textit{V}-band magnitudes. The \textit{K}-magnitudes appear nearly constant, showing little to no variation over time, whereas the \textit{V}-magnitudes exhibit a larger spread or dips lasting for months, likely caused by active pulsations and variable dust \citep{Kissetal.2006, Masseyetal.2009}. This behavior can be attributed to lower bolometric corrections and minor extinction values in the \textit{K}-band photometric region \citep{CCM.89}.

In light of the calibration, we followed the previous steps that conducted \textit{P}-\textit{L} relation with the use of \textit{K}-magnitudes \citep[e.g.][]{Taburetal.2009,Chatysetal.2019}.

To collect the required dataset, we sourced the available NIR databases; Catalogue of Infrared Observations \cite[CIO;][]{CIO.1999}, Catalogue of Stellar Photometry in Johnson's 11 Color System \cite{CSPCS.2002}, COBE DIBRE Point Source Catalog and Near-Infrared Light Curves \cite[DIBRE PSC and NIR LC;][]{DIBREPSC.2004,NIRLC.2010}, and the Two Micron All-Sky Survey \cite[2MASS;][]{2MASSPSC.2003}. Things to note here are that the \textit{K}-band magnitudes are taken on a single (nightly) observations and only 2MASS mission slightly differs from the rest by using a 'short-burst' duration of \textit{K}-band photometry (known as $K_S$-band) while providing a measurement of error. With that, we attempted to convert this photometric observation into a \textit{K}-bandpass to combine and determine the light source of interest's mean NIR magnitude. We calculated $\alpha$ Ori's \textit{L} provided by \citet{Daviesetal.2013} and their linear-fitted empirical relation extracted from RSGs' spectral energy distribution (SED) which corresponds to: 

\begin{equation}
\text{log}(L/L_{\odot}) = \alpha + \beta(m_\lambda - \mu)
\end{equation}

Where $\alpha$ and $\beta$ are the applicable parameters for most of RSGs, $m_\lambda$ is the apparent brightness subtracted for extinction at a particular wavelength \citep[\textit{K}-band for this study; see Table 4 for appropriate coefficients of][]{Daviesetal.2013}, and $\mu$ is the distance modulus (5log $d - 5$). A potential source of systematic error is the accounting for extinction caused by both circumstellar and interstellar foregrounds \citep{Masseyetal.2005,deWitetal.2008,Mattilaetal.2012}. However, the circumstellar material is not fully thick, allowing the optical bandpass to penetrate. This results in only a small effect ($A_V \leqslant 1.0$) as photons are scattered by embedded dust grains within the line of sight \citep{Kochaneketal.2012}, indicative of minor corrections are required in the IR region. The $\alpha$ and $\beta$ parameters' uncertainties, including the luminosity distance $\mu$, are the other sources of error that could also generate an offset at $\sim$0.1 dex and $\sim$0.2 dex in the IR and optical area, respectively. Therefore, if the extinction coefficient can be estimated and corrected in the process, \textit{K}-mags can measure the \textit{L}\textsubscript{bol} of an RSG, as well as possibly applicable to other photometric flavors.
\section{Results and Discussion}   

\subsection{Period Analysis}

\begin{figure}[ht!]
\begin{center}
\includegraphics[width=\columnwidth]{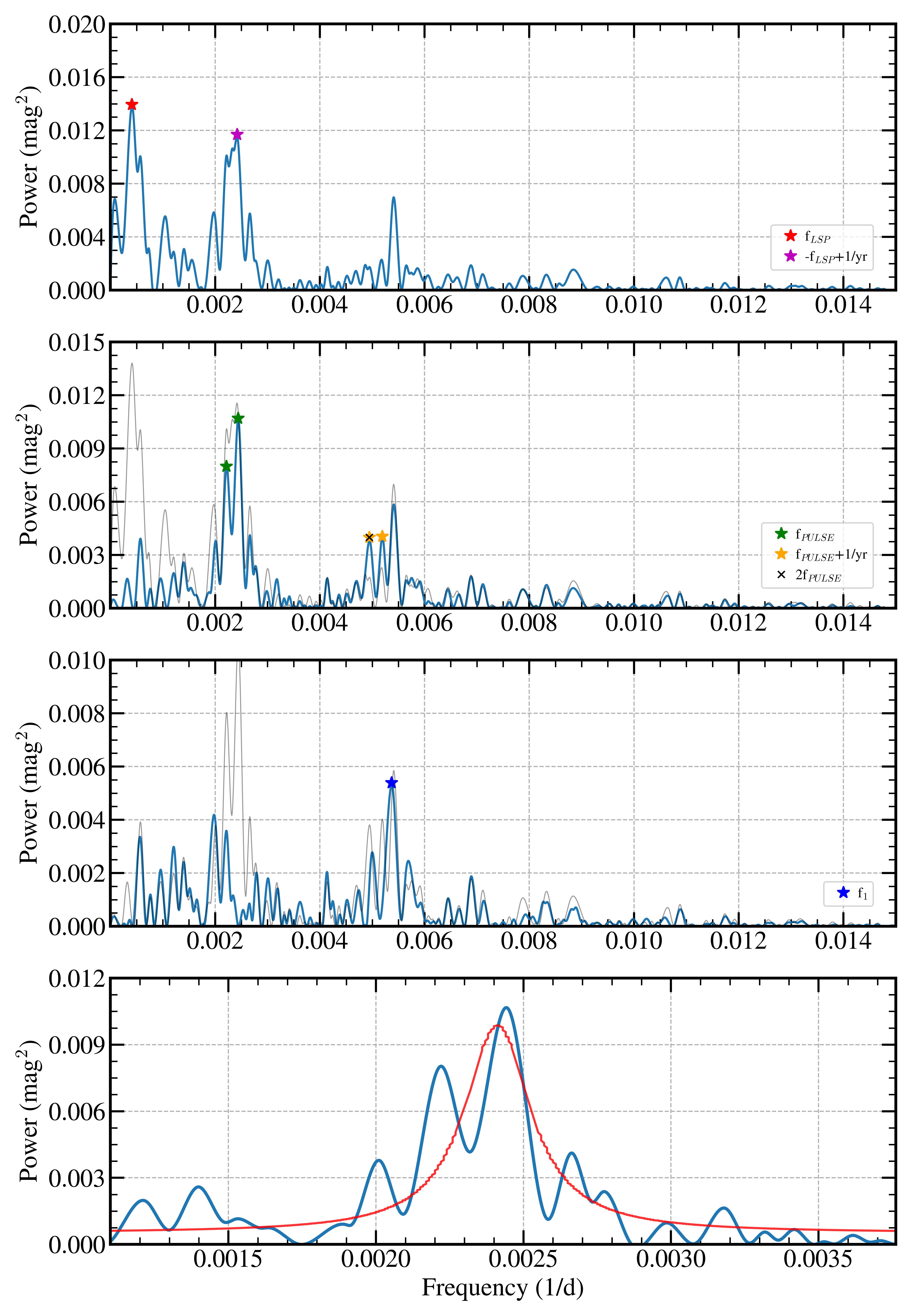}
\end{center}
\caption{Power spectra of $\alpha$ Ori's photometry. \textit{1st Column}: The strongest LSP and yearly aliases at its peak. \textit{2nd Column}: Deduced from the $f_{\text{LSP}}$ is the FM radial pulsation and secondary aliases labeled accordingly. \textit{3rd Column}: The remains of spectrum after the initial and second frequency extraction while standing tall is being marked as $f_{1}$ component. \textit{4th Column}: Lorentzian stochastic fitting along the $f_{\text{pulse}}$ spectrum.}
\label{fig:periodfreq}
\end{figure}

Initially, we are able to distinguish the extended periodicity which is equal to $f_{\text{LSP}} = 0.0004254$ d$^{-1}$ or \textit{P}$_{\text{LSP}}$ = 2350 $\pm$ 10 d. This is considerably longer to \citet{Kissetal.2006}. Meanwhile, the placement of this long period stands at the middle ground of those from the results of \citet{Kafka.2020} and \citet{Joyceetal.2020}. As noted by the previously mentioned authors and seen in Figure \ref{fig:periodfreq}, the $-f_{\text{LSP}}+1/yr$ conceals the sign of the dominant pulsation indicative of the point source's strong yearly aliases. 

We detect the split pulsation components $f_{\text{pulse1}} = 0.002445 $ d$^{-1}$ and $f_{\text{pulse2}} = 0.002013 $ d$^{-1}$. Two harmonics were then consolidated, fitted, and refined to determine the central dominant signature. Subsequently, we find a strong peak and calculated the fundamental mode (FM) of pulsation to be $f_{\text{pulse}} = 0.002410 $ $\pm$ 0.000176 d$^{-1}$ (\textit{P}$_{\text{0}}$ = 415 d $\pm$ 30 d). Moreover, it is well known that acoustic modes largely depends on the radius i.e. the average density of the star. This is known as the Period-Radius ($P$-$R$) relationship. Calculated using the \textit{Wesselink}
technique, the radial pulsation constant (which we labelled as $Q_{RP}$)\footnote{Practical application is developed by \citet{Balona.1979} used to determine the radius of a star or amplitude of the harmonic oscillation.} for cool stars turns out to be approximately 2 -- 4 \citep{Balona.1979,Lovyetal.1984}. From the equation below, the radial velocity amplitude of the FM takes into account the relationship between the star's radius and FM harmonics:

\begin{equation}
    Q_{RP} = \frac{A_{RV}}{A_{V}} \times \frac{P_{0}}{R_{0}}
\end{equation}
\begin{equation}
    A_{RV} = Q_{RP} \times A_{V} \times \frac{P_{0}}{R_{0}} \label{eq3}
\end{equation}

\begin{figure}[ht!]
\begin{center}
\includegraphics[width=\columnwidth]{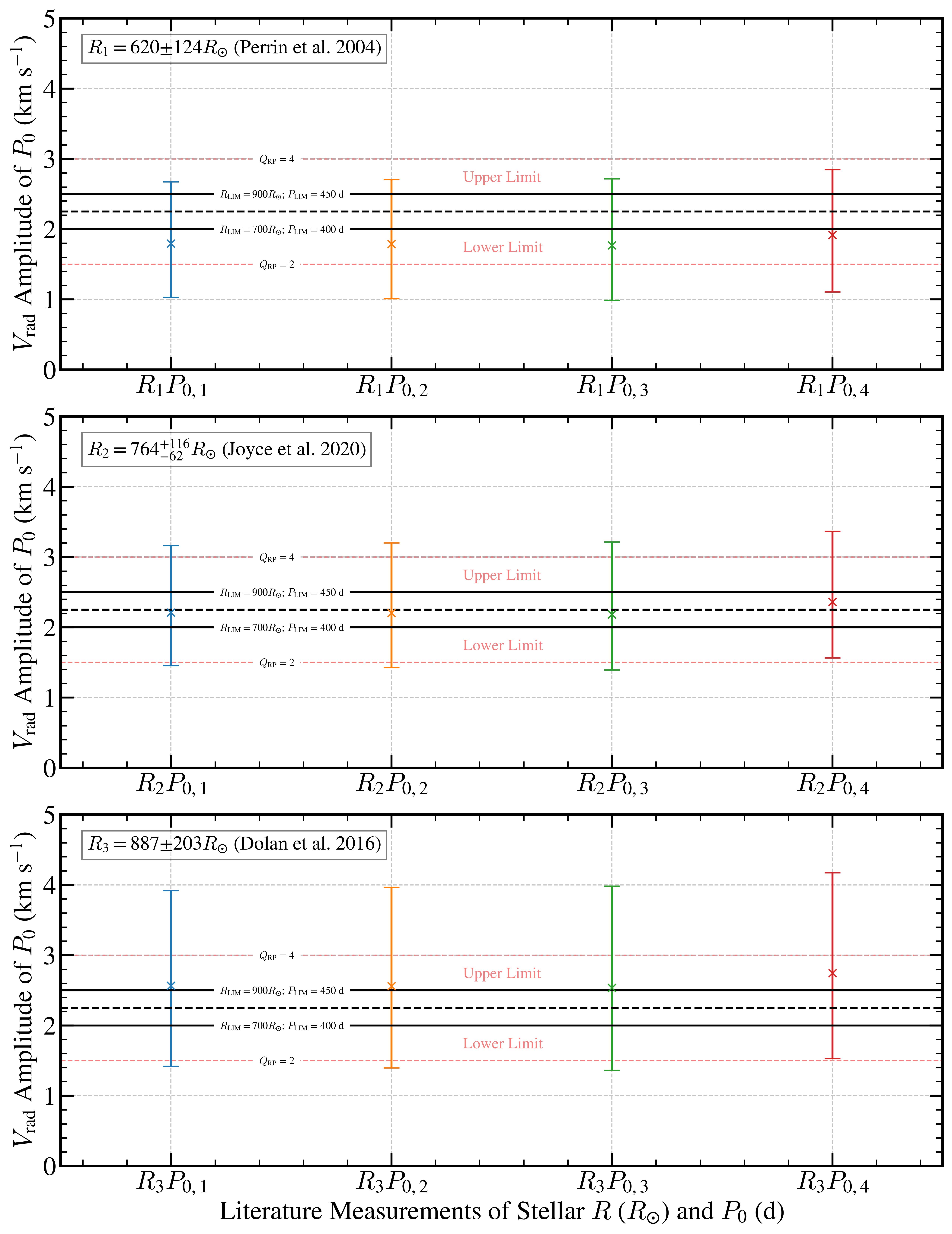}
\end{center}
\caption{Radial Velocity Amplitudes of FM harmonics based on a reference Stellar $R$ and $P_{\text{0}}$. \textit{1st Column}: \textit{R}\textsubscript{1} = 620 $\pm$ 124 $R_{\odot}$ \citep{Perrinetal.2004}. \textit{2nd Column}: \textit{R}\textsubscript{2} = $764_{-62}^{+116}$ $R_{\odot}$ \citep{Joyceetal.2020} \textit{3rd Column}: \textit{R}\textsubscript{3} = 887 $\pm$ 203 $R_{\odot}$ \citep{Dolanetal.2016}. Reference FM pulsation is as follows: $P_{\text{0,1}}$ = 415 $\pm$ 30 d (blue; this work), $P_{\text{0,2}}$ = 416 $\pm$ 24 d \citep[yellow; ][]{Joyceetal.2020}, $P_{\text{0,3}}$ = 420 $\pm$ 18 d \citep[green; ][]{Dupreeetal.1987}, $P_{\text{0,4}}$ = 388 $\pm$ 30 d \citep[red; ][]{Kissetal.2006}. The adjusted and projected range of $A_{RV}$ is depicted as black solid lines, with corresponding median (dashed black line), and both upper and lower limits (dashed red lines).}
\label{fig:amplitude}
\end{figure}

Inserting quantities for $\alpha$ Ori ($Q_{RP}$ = 3 $\pm$ 1; $A_{V}$ = 0.4 mag; \textit{R}$_{\text{0}}$ = $764_{-62}^{+116}$ $R_{\odot}$; \textit{P}$_{\text{0}}$ = 415 $\pm$ 30 d)\footnote{$Q_{RP}$ value is median quantity from \citet{Lovyetal.1984}, adopted radius is from the seismic analysis of \citet{Joyceetal.2020}, and the amplitude of the \textit{V}-band mag is estimated since there is no significant wavelength dependence in variation of \textit{B}, \textit{V}, and \textit{R}-bands \citep{Guin.1984}. Do not confuse it with the extinction coefficient.}, we predict the radial velocity amplitude of the FM pulse to be $2.21_{-0.50}^{+0.95}$ km s$^{-1}$ which is still confined (at least in the uncertainties) in the limits projected by \citet{Dupreeetal.1987}. However, the calculation from the aforementioned work assumes a smaller stellar radius, while the one adopted in this study is based on a larger stellar radius derived from linear perturbations. Using the available radius measurements, we compared and rectified the $A_{RV}$ limits presented by the previous authors. Figure \ref{fig:amplitude} illustrates the comparison of $\alpha$ Ori's known stellar radii and FM pulsation periods from multiple references. For a strict radius range of $\sim$700 – 900 $R_{\odot}$ and an FM pulsation period of $\sim$400 – 450 d, the radial velocity amplitude of the FM pulsation was adjusted to approximately 2.0 – 2.5 km s$^{-1}$, based on analyzing multiple combinations of stellar radius and FM pulsation, with upper and lower bounds of $\pm$ 0.5 km s$^{-1}$ permitted by $\sigma$$Q_{RP}$. 

Computed using Equation \ref{eq3}, the only $A_{RV}$ that coincides with this strict projection is the one where the seismic radius was adopted, while the other two stellar radii of $\alpha$ Ori remain within the upper and lower limits of the adjusted range. Moreover, the uncertainties in each case remain loose relative to the $A_{RV}$ range limits, primarily due to uncertainties in the radial pulsation constant and radius measurements, such as the 887 $R_{\odot}$ value reported by \citet{Dolanetal.2016}. Notably, a careful analysis reveals that if a radius of 887 $R_{\odot}$ is used, the period of $\alpha$ Ori should be 500 $\pm$ 40 d \citep{Joyceetal.2020}, rather than $\leqslant$ 450 d. This would result in an $A_{RV}$ of approximately 1.7 km s$^{-1}$, comparable to the result obtained when a radius of 620 $R_{\odot}$ is assumed. Hence, addressing the radial pulsation constant applicable to RSGs and obtaining a precise radius measurement — such as the seismic radius adopted for $\alpha$ Ori — can help further constrain and reduce the uncertainties of this practical parameter. Interestingly, the amplitude variation range presented in this work, compared to measurements reported both before the GDE (5 – 6 km s$^{-1}$) and during the GDE \citep[10 km s$^{-1}$;][]{Harper_2020,Dupreeetal.2020}, is significantly lower. This suggests that the GDE was a dynamic photospheric event that arose from a more vigorous pulsation episode.

Using the Full Width-Half Maxima method; FWHM, under the Lorentzian function ($\Gamma$) of $\tau$ = $\frac{1}{\pi\Gamma}$ to account for the internal stochastic signals, the cyclicity or mode lifetime of the point source can span for 1162 d ($\simeq$ 3 pulsation cycles). The acquired damping time matches well with the derived oscillating envelope of \citet{Kissetal.2006} and \citet{Joyceetal.2020}. Correspondingly, the number of pulsation in days is still on tight with \citet{Dupreeetal.1987} and their 3 year derived $\sim$ 420 d while it lags behind on the seismic FM pulsation only by a single day.

In addition, the single internal node known as first overtone (O1) component is also detected with $f_{\text{1}} = 0.00540$ $\pm$ 0.0003 d\textsuperscript{--1} (\textit{P}$_{\text{1}}$ = 185 d). The notable difference among recent cycle analysis for this star, however, is that the seedling of O$_1$ component stayed distinct during 1st and 2nd extraction at $\sim$0.0070 mag and 0.0058 mag, respectively, greater than the 2$f_{\text{pulse}}$ and $f_{\text{pulse}}+1/yr$ signatures. We, thus, identified a period ratio of \textit{P}$_{\text{1}}$/\textit{P}$_{\text{0}}$ = 0.446 which is also comparable to \citet{Joyceetal.2020}. As noted by previous authors (and references therein), pulsation models suggest that the periodic fraction for RGs and RSG pulsators should play at \textit{P}$_{\text{1}}$/\textit{P}$_{\text{0}}$ $\simeq$ 0.5 yet the lower mass ranges are the focus of such initiatives and likely would differ to higher regime.

\subsection{Bolometric Luminosity Calibration}

$\alpha$ Ori lies outside the Angelfish Nebula and the quadrangle of the $\lambda$ Ori (Meissa, HD 36861) emission cloud, aligning instead with the Ori-Eri Superbubble. In this region, foreground extinction and an excessive reddening law likely contribute to measurement errors along the line of sight. We propose that the encapsulated multiple dust shells in the visual regime still induce absorption, with silicate grain sizes around $\sim$0.1 $\mu$m. To outweigh this, \textit{K}-band photometry is essential for examining RSGs shrouded in interstellar material or dusty environments due to M\textsubscript{L}. Minor corrections and uncertainties are introduced during the de-reddening process, where the excess index $E_{(B-V)}$ is converted to extinction at a specific wavelength $A_{\lambda}$ \citep{CCM.89}.

To determine the colors, we inspected and approximated the color reddening of $\alpha$ Ori. Most of the inherent color values examined fall within the range $(B-V)$ = 1.80 – 1.89 \citep[e.g.][]{Mermilliod.1987,Mermilliod.2006}. From the extracted spectra, the flux and synthetic photometry show random errors of 3\% or better in most cases. We assigned a median intrinsic value of $(B-V)$ = 1.85, positioned near the midpoint of the extremes reported by \citet{Nicolet.1978} and \citet{HoffWarr.1991}, to align with the typical residual color of RSGs at $E_{(B-V)} \geqslant 0.3$ mag, and thus, its NIR extinction later on \citep[as a fixed value, e.g.][]{DaviesBeasor.2018}.

One may also explore combining optical and NIR colors, such as a $(V-K)$ diagram, which offers clearer measurements, particularly for RSGs like $\alpha$ Ori \citep[see,][]{Masseyetal.2009, Levesque.2018}. However, conventional color indices are sufficient for this analysis. After gathering the synthetic photometry, we focus on determining the spectral color index of interest, $(B-V)_0$. To aid in this, \citet{Fitzgerald.1970} provided a table associating $(B-V)_0$ values with the spectral types of M supergiants, assigning specific color indices to corresponding spectral classes.

\begin{equation}
E_{(B-V)} = (B-V) - 1.65
\end{equation}

Where $(B-V)_0$ = 1.65 as the tabulated value of an M2Iab RSG. As we further scan in various databases, there are few existing selections of $E_{(B-V)}$ for $\alpha$ Ori which are seldom used \citep[e.g. CPIRSS and ASAS-SN;][]{HindHarr.1994,ASASSN.2014,ASASSN.2019}. We make use of these residual colors and calculate the mean excess color index to achieve uniformity. With an average and fixed value of $E_{(B-V)}$ = 0.340, the selected NIR photometry is reddened using the \textit{K}-extinction factor of $R_K$ = 0.382 relative to canonical reddening value of $R_V$ = 3.1 \citep{CCM.89}\footnote{Although we do not have any knowledge of the appropriate relative law value $R_K$ (and the 'standard' $R_V$) applicable to that of $\alpha$ Ori's dusty environment, we rather assume that the extinction curve is similar to a diffuse background in the MW \citep[e.g.][]{Fitzpatrick.2004} and so as a decent choice for the object of interest at the time being.}:  

\begin{table*}[h!t!]
    \centering
    \begin{tabular}{lllll}
    \hline\hline
    Color Legend & Period $P_{\text{0}}$ & Luminosity \textit{L} & Notes & References  \\
    \hline
    Red star & 388 $\pm$ 30 d & $4.74_{-0.26}^{+0.31}$ & & \cite{Kissetal.2006,Dycketal.1992}  \\
    Orange star & 400 d* & 4.80 $\pm$ 0.19 & Proxy $P_{\text{0}}$ and No $\sigma$$P_{\text{0}}$ & \cite{Perrinetal.2004}  \\
    Green star & 415 $\pm$ 30 d & 5.00 $\pm$ 0.15 & & this work \\
    Teal star & 416 $\pm$ 24 d & $4.91_{-0.22}^{+0.19}$ & & \cite{Joyceetal.2020,Chatysetal.2019} \\
    Blue star & 420 $\pm$ 18 d & $4.94_{-0.06}^{+0.10}$ & $P_{\text{0}}$ = 1.15 $\pm$ 0.05 yr & \cite{Dupreeetal.1987,Joyceetal.2020} \\
    Orchid star & 435 d* & $5.10_{-0.21}^{+0.19}$ & Proxy $P_{\text{0}}$ and No $\sigma$$P_{\text{0}}$ & \cite{Harperetal.2008} \\
    \hline\hline
    Name & Period $P_{\text{0}}$ & Luminosity \textit{L} & Notes & References \\
    \hline
    SRc SU Per & 470 $\pm$ 70 d & $5.36 \pm 0.23_{(-0.71)}^{(+0.67)}$ & & \citet{Kissetal.2006,Chatysetal.2019} \\
    SRc RT Car & 435 d & $5.09_{-0.21(-0.54)}^{+0.20(+0.97)}$ & No $\sigma$$P_{\text{0}}$ & \citet{Chatysetal.2019} \\
    SRc IX Car & 408 $\pm$ 50 d & $4.89_{-0.20(-0.63)}^{+0.18(+0.51)}$ & & \citet{Kissetal.2006,Chatysetal.2019} \\
    SRc YZ Per & 378 d & $4.67_{-0.18(-0.56)}^{+0.20(+0.61)}$ & No $\sigma$$P_{\text{0}}$ & \citet{Verhoelstetal.2009} \\
    \hline
    \end{tabular}
    \caption{Combinations and data log of $\alpha$ Ori and annotated Galactic RSGs for comparison adopted and used in Period-Luminosity Relationship. Annotated with asterisks are acting as arbitrary pulsation periods that are close to those of acquired. The 400 d is an median difference between 388 d and 415 d while, 435 d is an upper-limit permitted by uncertainty as in \cite{Guinan.2019}. Meanwhile, for the set of RSGs, a revised astrometric distance from \textit{Hipparcos} was used for YZ Per, while the rest are from \textit{Gaia} DR2. Included are the 3$\sigma$ limits in parentheses.}
    \label{table:1}
\end{table*}

\begin{equation}
    k_{\lambda} = \frac{A_\lambda}{E_{(B-V)}}
\end{equation}
\begin{equation}
    A_{K} = R_{K} E_{(B-V)}
\end{equation}

Where $A_K$ represents the absolute-to-selective extinction coefficient in the NIR \textit{K}-region \citep{Fitzpatrick.1999}. However, a notable caveat is that common extinction values are predominantly derived using main-sequence OB stars, which have spectral energy distributions (SEDs) distinct from those of red supergiants (RSGs) like $\alpha$ Ori. This distinction suggests that RSGs can appear redder in a given filter and effective wavelength when utilized. Consequently, slightly higher reddening values are required, such as $R_V \simeq 4.2$, in the classical passband calibrated for late-type and dusty RSGs \citep{Eliasetal.1985, Nakayaetal.2001, Masseyetal.2005}.

Onwards, despite the uncertainties embedded in each NIR point being variable by $\pm$ 0.2 mag, $\alpha$ Ori's \textit{K}-amplitude predominantly lies around $\sim$$-4$ mag, indicating a more stable value compared to conventional optical measurements under conditions of lower reddening. The chosen \textit{K}-magnitudes are the preferred band for calculating and estimating point sources' \textit{L}${\text{bol}}$ and its proxies (\textit{M}${\text{bol}}$/\textit{M}$_{\text{K}}$), as they are close to the flux peak, allowing absorption effects to be 'almost' negligible, except in cases where a few measurements are saturated or undersaturated (likely due to instrumental errors or the influence of an unforeseeable companion). We begin by converting the 2MASS $K_S$ photometry using the relationship provided by \citet{Carpenter.2001}:

\begin{equation}
    k = K_S + 0.04
\end{equation}

The converted $K_S$ magnitude is then mixed and processed to obtain the mean \textit{K}-band photometry alongside its standard error. We then proceed to subtract the extinction and perform calibration to \textit{L}. Derived from the mean \textit{K}-magnitude, the \textit{L}\textsubscript{bol} is known and applied for:

\begin{equation}
    m_k = k - 0.130
\end{equation}
\begin{equation}
    \text{log}(L/L_{\odot}) = 0.90 - 0.40(-4.130 - 6.127)
\end{equation}

As the \textit{K}-extinction $A_K$ = 0.130 and subsequently, $\mu$ = $6.127_{-0.20}^{+0.33}$ is derived from the present-day 3$\sigma$ seismic distance \citep{Joyceetal.2020}. Hence, the \textit{K}-band derived \textit{L} for $\alpha$ Ori equates to log($L$/$L_{\odot}$) = 5.00 $\pm$ $0.15_{(-0.45)}^{(+0.48)}$ at conventional 1$\sigma$ and 3$\sigma$ limits, respectively. 

In comparison, this result is smaller but within the uncertainty range of the \textit{L}\textsubscript{bol} obtained from radio measurements. While it is slightly higher by 0.6 mag, the value remains less constrained compared to the pronounced stellar \textit{L}, though still consistent with the imposed \textit{T}\textsubscript{eff} range. Furthermore, the \textit{K}-band extinction and color excess for $\alpha$ Ori from previous steps align well with the 'standard' values and recently derived corrections for the chosen bandpass applicable to most RSGs \citep{Schlegeletal.1998,DaviesBeasor.2018}. As a trade-off from the linear fitting approach, this suggests a higher effective temperature for the point source, indicating that $\alpha$ Ori may be hotter by approximately $\pm$ 200 K when using TiO-based measurements.

\subsection{Period-Luminosity: Diagnostic Approach}

\begin{figure*}[ht!]
\begin{center}
\includegraphics[scale=0.43]{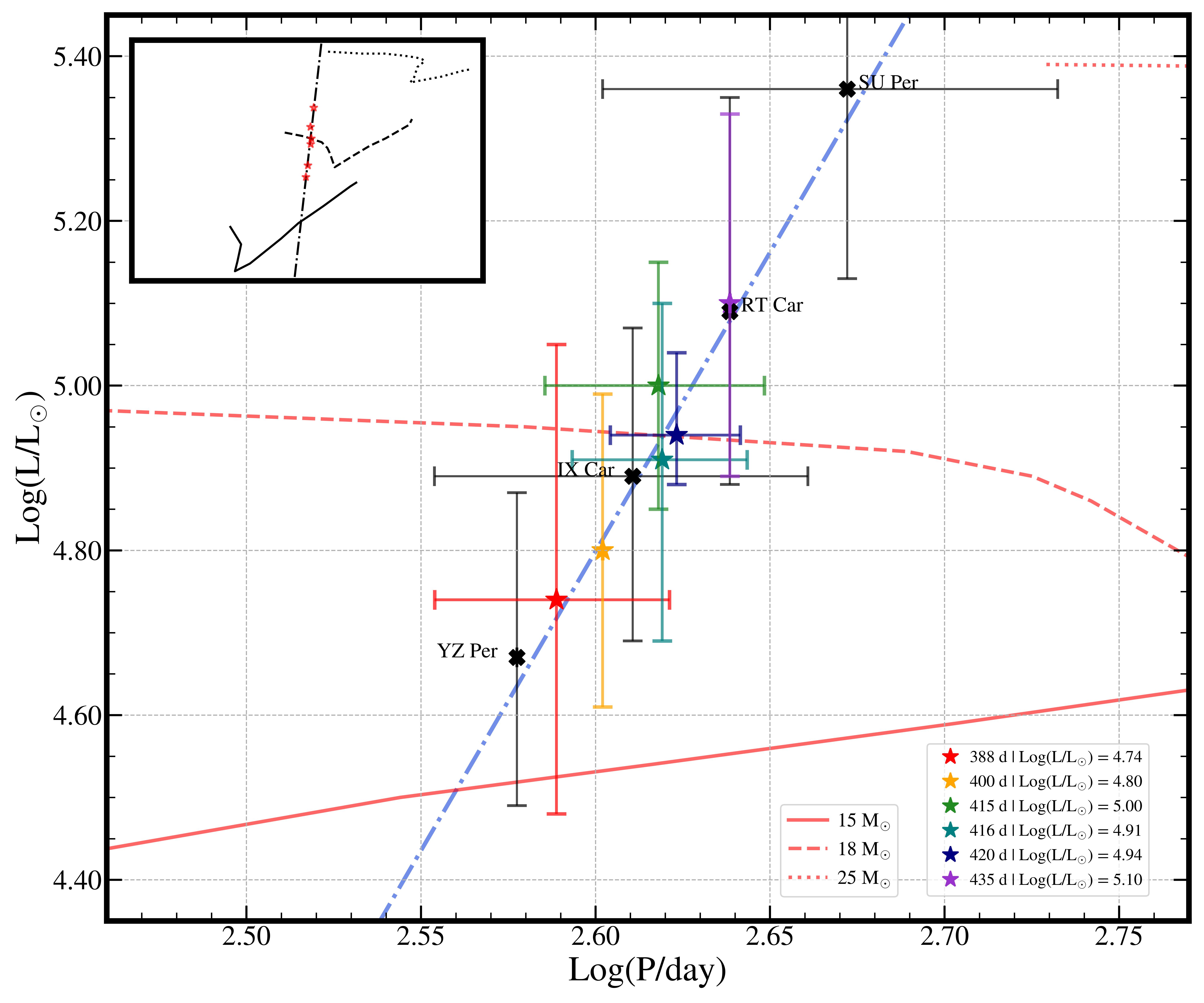}
\end{center}
\caption{Period-Luminosity diagram for $\alpha$ Ori. The colored stars denote the close-proximity harmonics aligned with an array of \textit{L} in order. The associated error bars are the variations given to each \textit{L} quantity, while only 4 of $\alpha$ Ori's $P_{\text{0}}$ have uncertainty limits. The red lines, on the other hand, are based on non-adiabatic initiatives of \cite{GuoLi.2002} on FM pulsation for 3 sets of stellar mass along a convective mixing-length parameter, $\alpha$\textsubscript{MLT} = 1.0 and metallicity, \textit{Z} = 0.02. Also added are cross marks representative of galactic samples with comparable semi-oscillations and \textit{L}$_{\text{bol}}$ values.}
\label{fig:plrelate}
\end{figure*}

Variable stars, especially Cepheids, are regularly called as 'Standard Candles' as they can be used to estimate the distance from a single point to another \citep{Leavitt.1912}. This method was later on applied to other pulsators as instruments and grasps got better ranging from RR-Lyrae stars concentrated in the middle of the Horizontal Branch (HB) up to large Long Period Variables/LPVs (\textit{Prototypes:} Miras and SRs\footnote{Semi-regular Variable Stars}), particularly for RGs and RSGs wherein their radial pulsation are caused by dramatic internal changes \citep{Catelanetal.2004,Aslan.2004}. Though the relation for SR-type RSGs are not quite tight, with their brightness they can also be used for distance ladder method in extragalactic manner.

For this part, we acquired six of the best observational measurements of luminosities and periods from previous works around $\alpha$ Ori to test Leavitt's Law (the $P$-$L$ relationship). However, there are some considerations when performing the best-fitting procedure. First, we transformed the 1$\sigma$ revised \textit{Hipparcos} parallax reported by \citet{Chatysetal.2019}, where the star's distance is close to the value determined by \citet{Lambertetal.1984}. Nevertheless, the astrometric parameters still contain some systematic errors due to the inclusion of cosmic noise. Following the same methodology as before, we calculated $\mu$ and utilized the same \textit{K}-magnitude with its extinction corrected to determine \textit{L}\textsubscript{bol}. Using the revised \textit{Hipparcos} solution yields log($L$/$L_{\odot}$) = $4.91_{-0.22(-0.79)}^{+0.19(+0.25)}$, where the initial error given are its 1$\sigma$ intervals, while the 3$\sigma$ cut-offs are in parentheses. This value remains close to that of \citet{Joyceetal.2020} and is merged with the radial period of 416 days, while the 3$\sigma$ stellar \textit{L} is combined with the 420 $\pm$ 18 days by \citet{Dupreeetal.1987} to act as a 'benchmark' for the fitting line. The standard procedure from prior calibrations is then applied to the remaining data, excluding the recently acquired values. Table \ref{table:1} summarizes all the matched data utilized for this analysis, while Figure \ref{fig:plrelate} shows the measurements' common trend line. A best-fitting line from the $P$-$L$ log space yields:

\begin{equation}
    \text{log}(L/L_{\odot}) = 7.26 \pm 0.10 \times \text{log \textit{P}} + (-14.10 \pm 0.25)
\end{equation}

Included in the previously mentioned \textit{P}-\textit{L} figure are the linear modelling calculations for dominant pulsation of 15 -- 25 $M_{\odot}$ Galactic RSGs with a fixed metal fraction \citep{GuoLi.2002}. Using the \textit{K}-band correction of \citet{DaviesBeasor.2018} of $A_K$ = 0.119, we also sampled a few selections of $\alpha$ Ori's galactic 'doppelgangers' that are RSGs from two stellar neighborhoods with fundamental overtones in like to those quantified periods of $\alpha$ Ori \citep[see,][]{Verhoelstetal.2009,Chatysetal.2019}.

Although non-adiabatically modelled and with lower mixing-length treatment, we can see that the 'benchmark' value (Blue star) almost converges on 18 $M_{\odot}$ mark (dashed). The \textit{L}\textsubscript{bol} of $\alpha$ Ori we achieved mainly contributes to the deviation due to the constricted distance but shows that $\alpha$ Ori can be situated into the lower bound 18 $M_{\odot}$ regime caused by the acquired pulsation. For comparison among the galactic samples (in terms of \textit{L} and $P_{\text{0}}$), IX Car is nearly closing onto the intersection line where the proximity of \textit{L} quantities and FM periods converge for $\alpha$ Ori. While YZ Per is likely underestimated due to the errors in distance but a semi-overtone close to \citet{Kissetal.2006}, SU Per registered a quasi-period approaching the outer limits of the $\sim$ 420 d on top of log($L$/$L_{\odot}$) $\sim$ 5.3 hierarchy. Further addressing the mentioned galactic pulsator on weighing to $\alpha$ Ori, the oscillation growth rates of SRs are strongly erratic accompanied with pronounced modulation \citep{Buchleretal.1996}. Such changes in pulsation patterns are accounted for in hydrodynamical models of $\alpha$ Ori at $\geqslant$ 19 $M_{\odot}$, where \textit{L}, \textit{T}\textsubscript{eff}, and periodicities are influenced by the $\kappa$-mechanism and other opacities. Thus, the RSG of interest may converge toward RT Car and SU Per's current \textit{L} and quasi-oscillation as it evolves over its lifetime.

\section{Conclusion and Recommendation}

We have presented a concise overview of $\alpha$ Ori's photometric variation, an approximate value for \textit{L}\textsubscript{bol}, and a benchmarking tool for $\alpha$ Ori. We determine a long secondary period of \textit{P}$_{\text{LSP}}$ = 2350 $\pm$ 10 days, which lies within the middle and in range of recent literature findings. Likewise, an FM pulsation of \textit{P}${\text{0}}$ = 415 d $\pm$ 30 d, with an estimated radial velocity amplitude of $A_{RV}$ = $2.21_{-0.50}^{+0.95}$ km s\textsuperscript{--1} (adopted seismic radius), was determined, supporting the current literature on this subject. An adjustment to the previously projected $A_{RV}$ range was performed, yielding a revised range of approximately 2.0 – 2.5 km s\textsuperscript{--1}, derived using multiple factors: (1) stellar radius measurements from previous studies, (2) FM pulsation data from this work and other cited sources, and (3) applying constraints of a stellar radius range of $\sim$700 – 900 $R_{\odot}$ and FM harmonics between $\sim$400 – 450 d. Lastly, a signature of 185 d mode was identified later on as first overtone. Although, we cannot fully consider that the $f_{\text{1}}$ component does exist in our periodogram extraction due to photometric gaps, we can rather end up that $\alpha$ Ori likely has second short-periodic shift supporting the notion of previously-mentioned proponents. More photometric observations and thorought light curve analysis are required for this RSG to validate the essence of the 185 d component. 

In addition, we conducted an approximation of $\alpha$ Ori’s \textit{L}\textsubscript{bol} using 3$\sigma$ seismic parallax of \citet{Joyceetal.2020} to achieve a distance modulus $\mu$ = $6.127_{-0.20}^{+0.33}$. We, then, sourced for \textit{K}-magnitudes in which we averaged out across the board and subtracted in extinction. This luminosity approximation led us to a log($L$/$L_{\odot}$) = 5.00 $\pm$ $0.15_{(-0.45)}^{(+0.48)}$. The statistical result, meanwhile, for Leavitt's \textit{P}-\textit{L} diagram is quite expected ($R^2$ $\sim$ 0.91). Whereas a far more reasonable approach is to gather a selection of RSGs to truly appreciate the relation \citep[e.g.][]{Chatysetal.2019}. However, it quite shows to us that $\alpha$ Ori can be situated at around 18 $M_{\odot}$ range using the combined best observational results of seismic luminosity log($L$/$L_{\odot}$) = $4.94_{-0.06}^{+0.10}$ and fundamental pulsation $P_{\text{0}}$ = 420 $\pm$ 18 d. Key limitation of this benchmark result however is that the stellar models of \citet{GuoLi.2002} are done non-adiabatically with lower mixing-length treatments. This can produce stellar models with inappropriately low effective temperatures that are sub-solar in nature \citep{LevesqueMessey.2020}. Whereas more contemporary techniques supply $\alpha$\textsubscript{MLT} $\geqslant$ 2.0 and adiabatic schemes for linear seismic analysis \citep[e.g.][and references therein]{Joyceetal.2020,Goldberg_2022}\footnote{Although \citet{Chatzopoulosetal.2020} post-merger calculations in light of $\alpha$ Ori's rotation utilized a slightly lower $\alpha$\textsubscript{MLT} formalism.}.

$\alpha$ Ori lies within the upper convective part of the HRD. It is no secret that the regularity of the photometric variations of the SRc-type RSG are likely to be caused and affected by the radial FM pulsations and inherently by half the stellar radius-size granulation cells, flows, and hot spots bringing excitation and pressure, stochastically (or $\rho$-mode pulse) which also happens to harbor a dynamo action \citep{StothersLeung.1971,Schwarzschild.1975,Auriereetal.2010,Auriereetal.2016}. The latter features were previously theorized among RGs to luminous RSGs and was later on resolved and imaged using polarized spectra \citep{DorchFreytag.2003,LopezAristeetal.2018}. 

This radial pulsation is expected to scale upwards together with \textit{L} and more effectively towards the poleward axes of the RSG as the mentioned-star continues to evolve \citep{Dupree.2011}. As such, this may also give rise to multiple asymmetries from the accompanied periodicity which predicts the existence of longer periods (attributed to g-mode, convective cell turnover, or binarity etc.), O\textsubscript{1}, and possibly O\textsubscript{2} period oscillation. In fact, RGs and RSGs entailed with more than one stellar pulse are quite common in most, if not all, of the observed ones in the galactic plane. More photometric observations and on-through light curve analysis are required for this RSG to further validate the 185 d variation mode.

Although we have not touched modelling frameworks yet (e.g. MESA etc.) and the parallelism between present-day mass and pulsation trends are still grey caused by the physics of convective parameters, we consider that adiabatic systems with higher progenitor mass and same projection for its present-day mass can also induce a log($L$/$L_{\odot}$) $\sim$ 4.9 -- 5.0 along their $\sim$ 400 d semi-regular pulsation (and other well-matched constraints). This is precisely demonstrated for RSGs like $\alpha$ Ori \citep{YoonCantiello.2010,Goldbergetal.2020,Joyceetal.2020}\footnote{Not consistently held as \citet{Goldbergetal.2020} yielded a lower mass RSG but with high radius and stellar \textit{L}.}. One may further explore this in the future where one consolidates the observation and theoretical calculations e.g. stellar evolution models. This can be with the inclusion of rotation, pulsation profiles, and even binary/merger scenarios since majority of the descendants of RSGs are born in binary systems \citep{deMinketal.2013,deMinketal.2014,Chatzopoulosetal.2020}, to delineate the mass ranges to investigate around and other significant stellar parameters that require further rectification, especially with the new results on the binary companion of $\alpha$ Ori. 

\section*{Acknowledgments}

We sincerely appreciate the valuable comments of the three anonymous reviewers, and both the sector and managing editor of OjAp, which helped to improve this manuscript. CGHA and WADS would like to thank Dr. Emmanouil Chatzopoulos and Dr. Meridith Joyce for the hefty discussions regarding MESA to be implemented on a future work in conjunction and Dr. Moln{\'a}r L{\'a}szl{\'o} for the support and expert guidance in accordance to frequency analysis. Furthermore, the corresponding author would like to extend its gratitude to Prof. Willie Anthony D. Sapalaran (WADS) on aiding for this paper. CGHA also thanks himself for surviving and completing his undergraduate years. Substantial help was also provided by AAVSO, SMEI, and NIR catalogues used in this study. This brief research report has made use of the SIMBAD database, operated at CDS, Strasbourg, France, and NASA's Astrophysics Data System Bibliographic Services. \\
\\
\textit{Softwares:} AstroPy \citep{Astropy2022}, Matplotlib \citep{Matplotlib2007}, NumPy \citep{Numpy2020}, \href{https://www.originlab.com/}{OriginLab}, Pandas \citep{Pandas2011}, Period04 \citep{LenzBreger.2005}, SciPy \citep{Scipy2020}, SOAD \citep{SOAD2019} 

\section*{Conflicts of Interest}

The authors declare that they have no competing interests.

\section*{Data Availability Statement}

For the \textit{V}-band magnitudes, the observations can be found in \href{https://www.aavso.org/}{AAVSO} and its supplementary SMEI photometry is provided by Dr. Meridith Joyce and her team's work (see \href{https://arxiv.org/abs/2006.09837}{arXiv then \textit{comments}}) as text file. Meanwhile, the analyzed \textit{K}-band datasets that are in NIR Catalogues i.e. CIO, Johnson's 11 Color System, COBRE DIBRE PSC and NIR LC, and 2MASS can be found and queried in via \href{https://vizier.cds.unistra.fr/viz-bin/VizieR}{VizieR} operated by CDS. Other data and parameters were adopted through various literatures \citep[e.g.][]{GuoLi.2002,Dolanetal.2016,Chatysetal.2019,Joyceetal.2020}.


\newpage

\bibliography{bibfile}

\begin{thebibliography}{}
\expandafter\ifx\csname natexlab\endcsname\relax\def\natexlab#1{#1}\fi
\providecommand{\url}[1]{\href{#1}{#1}}
\providecommand{\dodoi}[1]{doi:~\href{http://doi.org/#1}{\nolinkurl{#1}}}
\providecommand{\doeprint}[1]{\href{http://ascl.net/#1}{\nolinkurl{http://ascl.net/#1}}}
\providecommand{\doarXiv}[1]{\href{https://arxiv.org/abs/#1}{\nolinkurl{https://arxiv.org/abs/#1}}}

\bibitem[{{Astropy Collaboration} {et~al.}(2022){Astropy Collaboration}, {Price-Whelan}, {Lim}, {Earl}, {Starkman}, {Bradley}, {Shupe}, {Patil}, {Corrales}, {Brasseur}, {N{\"o}the}, {Donath}, {Tollerud}, {Morris}, {Ginsburg}, {Vaher}, {Weaver}, {Tocknell}, {Jamieson}, {van Kerkwijk}, {Robitaille}, {Merry}, {Bachetti}, {G{\"u}nther}, {Aldcroft}, {Alvarado-Montes}, {Archibald}, {B{\'o}di}, {Bapat}, {Barentsen}, {Baz{\'a}n}, {Biswas}, {Boquien}, {Burke}, {Cara}, {Cara}, {Conroy}, {Conseil}, {Craig}, {Cross}, {Cruz}, {D'Eugenio}, {Dencheva}, {Devillepoix}, {Dietrich}, {Eigenbrot}, {Erben}, {Ferreira}, {Foreman-Mackey}, {Fox}, {Freij}, {Garg}, {Geda}, {Glattly}, {Gondhalekar}, {Gordon}, {Grant}, {Greenfield}, {Groener}, {Guest}, {Gurovich}, {Handberg}, {Hart}, {Hatfield-Dodds}, {Homeier}, {Hosseinzadeh}, {Jenness}, {Jones}, {Joseph}, {Kalmbach}, {Karamehmetoglu}, {Ka{\l}uszy{\'n}ski}, {Kelley}, {Kern}, {Kerzendorf}, {Koch}, {Kulumani}, {Lee}, {Ly}, {Ma}, {MacBride}, {Maljaars}, {Muna}, {Murphy}, {Norman},
  {O'Steen}, {Oman}, {Pacifici}, {Pascual}, {Pascual-Granado}, {Patil}, {Perren}, {Pickering}, {Rastogi}, {Roulston}, {Ryan}, {Rykoff}, {Sabater}, {Sakurikar}, {Salgado}, {Sanghi}, {Saunders}, {Savchenko}, {Schwardt}, {Seifert-Eckert}, {Shih}, {Jain}, {Shukla}, {Sick}, {Simpson}, {Singanamalla}, {Singer}, {Singhal}, {Sinha}, {Sip{\H{o}}cz}, {Spitler}, {Stansby}, {Streicher}, {{\v{S}}umak}, {Swinbank}, {Taranu}, {Tewary}, {Tremblay}, {de Val-Borro}, {Van Kooten}, {Vasovi{\'c}}, {Verma}, {de Miranda Cardoso}, {Williams}, {Wilson}, {Winkel}, {Wood-Vasey}, {Xue}, {Yoachim}, {Zhang}, {Zonca}, \& {Astropy Project Contributors}}]{Astropy2022}
{Astropy Collaboration}, {Price-Whelan}, A.~M., {Lim}, P.~L., {et~al.} 2022, \apj, 935, 167, \dodoi{10.3847/1538-4357/ac7c74}

\bibitem[{{Auri{\`e}re} {et~al.}(2010){Auri{\`e}re}, {Donati}, {Konstantinova-Antova}, {Perrin}, {Petit}, \& {Roudier}}]{Auriereetal.2010}
{Auri{\`e}re}, M., {Donati}, J.~F., {Konstantinova-Antova}, R., {et~al.} 2010, Astronomy and Astrophysics, 516, L2, \dodoi{10.1051/0004-6361/201014925}

\bibitem[{{Auri{\`e}re} {et~al.}(2016){Auri{\`e}re}, {L{\'o}pez Ariste}, {Mathias}, {L{\`e}bre}, {Josselin}, {Montarg{\`e}s}, {Petit}, {Chiavassa}, {Paletou}, {Fabas}, {Konstantinova-Antova}, {Donati}, {Grunhut}, {Wade}, {Herpin}, {Kervella}, {Perrin}, \& {Tessore}}]{Auriereetal.2016}
{Auri{\`e}re}, M., {L{\'o}pez Ariste}, A., {Mathias}, P., {et~al.} 2016, Astronomy and Astrophysics, 591, A119, \dodoi{10.1051/0004-6361/201628077}

\bibitem[{Balona \& Stobie(1979)}]{Balona.1979}
Balona, L.~A., \& Stobie, R.~S. 1979, Monthly Notices of the Royal Astronomical Society, 189, 649, \dodoi{10.1093/mnras/189.4.649}

\bibitem[{{Buchler} {et~al.}(1996){Buchler}, {Kollath}, {Serre}, \& {Mattei}}]{Buchleretal.1996}
{Buchler}, J.~R., {Kollath}, Z., {Serre}, T., \& {Mattei}, J. 1996, Astrophysical Journal, 462, 489, \dodoi{10.1086/177167}

\bibitem[{Buffington {et~al.}(2007)Buffington, Morrill, Hick, Howard, Jackson, \& Webb}]{Buffingtonetal.2007}
Buffington, A., Morrill, J.~S., Hick, P.~P., {et~al.} 2007, in Solar Physics and Space Weather Instrumentation II, ed. S.~Fineschi \& R.~A. Viereck, Vol. 6689, International Society for Optics and Photonics (SPIE), 79 -- 84, \dodoi{10.1117/12.734658}

\bibitem[{{Cardelli} {et~al.}(1989){Cardelli}, {Clayton}, \& {Mathis}}]{CCM.89}
{Cardelli}, J.~A., {Clayton}, G.~C., \& {Mathis}, J.~S. 1989, Astrophysical Journal, 345, 245, \dodoi{10.1086/167900}

\bibitem[{{Carpenter}(2001)}]{Carpenter.2001}
{Carpenter}, J.~M. 2001, Astronomical Journal, 121, 2851, \dodoi{10.1086/320383}

\bibitem[{{Catelan} {et~al.}(2004){Catelan}, {Pritzl}, \& {Smith}}]{Catelanetal.2004}
{Catelan}, M., {Pritzl}, B.~J., \& {Smith}, H.~A. 2004, Astrophysical Journal, 154, 633, \dodoi{10.1086/422916}

\bibitem[{{Chatys} {et~al.}(2019){Chatys}, {Bedding}, {Murphy}, {Kiss}, {Dobie}, \& {Grindlay}}]{Chatysetal.2019}
{Chatys}, F.~W., {Bedding}, T.~R., {Murphy}, S.~J., {et~al.} 2019, Monthly Notices of the Royal Astronomical Society, 487, 4832, \dodoi{10.1093/mnras/stz1584}

\bibitem[{{Chatzopoulos} {et~al.}(2020){Chatzopoulos}, {Frank}, {Marcello}, \& {Clayton}}]{Chatzopoulosetal.2020}
{Chatzopoulos}, E., {Frank}, J., {Marcello}, D.~C., \& {Clayton}, G.~C. 2020, Astrophysical Journal, 896, 50, \dodoi{10.3847/1538-4357/ab91bb}

\bibitem[{{Chiavassa} {et~al.}(2011){Chiavassa}, {Freytag}, {Masseron}, \& {Plez}}]{Chiavassaetal.2011}
{Chiavassa}, A., {Freytag}, B., {Masseron}, T., \& {Plez}, B. 2011, Astronomy and Astrophysics, 535, A22, \dodoi{10.1051/0004-6361/201117463}

\bibitem[{{Chiavassa} {et~al.}(2010){Chiavassa}, {Haubois}, {Young}, {Plez}, {Josselin}, {Perrin}, \& {Freytag}}]{Chiavassaetal.2010}
{Chiavassa}, A., {Haubois}, X., {Young}, J.~S., {et~al.} 2010, Astronomy and Astrophysics, 515, A12, \dodoi{10.1051/0004-6361/200913907}

\bibitem[{{Chiavassa} {et~al.}(2009){Chiavassa}, {Plez}, {Josselin}, \& {Freytag}}]{Chiavassaetal.2009}
{Chiavassa}, A., {Plez}, B., {Josselin}, E., \& {Freytag}, B. 2009, Astronomy and Astrophysics, 506, 1351, \dodoi{10.1051/0004-6361/200911780}

\bibitem[{{Cox} {et~al.}(2012){Cox}, {Kerschbaum}, {van Marle}, {Decin}, {Ladjal}, {Mayer}, {Groenewegen}, {van Eck}, {Royer}, {Ottensamer}, {Ueta}, {Jorissen}, {Mecina}, {Meliani}, {Luntzer}, {Blommaert}, {Posch}, {Vandenbussche}, \& {Waelkens}}]{Coxetal.2012}
{Cox}, N.~L.~J., {Kerschbaum}, F., {van Marle}, A.~J., {et~al.} 2012, Astronomy and Astrophysics, 537, A35, \dodoi{10.1051/0004-6361/201117910}

\bibitem[{{Cutri} {et~al.}(2003){Cutri}, {Skrutskie}, {van Dyk}, {Beichman}, {Carpenter}, {Chester}, {Cambresy}, {Evans}, {Fowler}, {Gizis}, {Howard}, {Huchra}, {Jarrett}, {Kopan}, {Kirkpatrick}, {Light}, {Marsh}, {McCallon}, {Schneider}, {Stiening}, {Sykes}, {Weinberg}, {Wheaton}, {Wheelock}, \& {Zacarias}}]{2MASSPSC.2003}
{Cutri}, R.~M., {Skrutskie}, M.~F., {van Dyk}, S., {et~al.} 2003, {2MASS All Sky Catalog of Point Sources}

\bibitem[{{Davies} \& {Beasor}(2018)}]{DaviesBeasor.2018}
{Davies}, B., \& {Beasor}, E.~R. 2018, Monthly Notices of the Royal Astronomical Society, 474, 2116, \dodoi{10.1093/mnras/stx2734}

\bibitem[{{Davies} {et~al.}(2013){Davies}, {Kudritzki}, {Plez}, {Trager}, {Lan{\c{c}}on}, {Gazak}, {Bergemann}, {Evans}, \& {Chiavassa}}]{Daviesetal.2013}
{Davies}, B., {Kudritzki}, R.-P., {Plez}, B., {et~al.} 2013, Astrophysical Journal, 767, 3, \dodoi{10.1088/0004-637X/767/1/3}

\bibitem[{{Davies} {et~al.}(2017){Davies}, {Kudritzki}, {Lardo}, {Bergemann}, {Beasor}, {Plez}, {Evans}, {Bastian}, \& {Patrick}}]{Daviesetal.2017}
{Davies}, B., {Kudritzki}, R.-P., {Lardo}, C., {et~al.} 2017, Astrophysical Journal, 847, 112, \dodoi{10.3847/1538-4357/aa89ed}

\bibitem[{{de Mink} {et~al.}(2013){de Mink}, {Langer}, {Izzard}, {Sana}, \& {de Koter}}]{deMinketal.2013}
{de Mink}, S.~E., {Langer}, N., {Izzard}, R.~G., {Sana}, H., \& {de Koter}, A. 2013, Astrophysical Journal, 764, 166, \dodoi{10.1088/0004-637X/764/2/166}

\bibitem[{{de Mink} {et~al.}(2014){de Mink}, {Sana}, {Langer}, {Izzard}, \& {Schneider}}]{deMinketal.2014}
{de Mink}, S.~E., {Sana}, H., {Langer}, N., {Izzard}, R.~G., \& {Schneider}, F.~R.~N. 2014, Astrophysical Journal, 782, 7, \dodoi{10.1088/0004-637X/782/1/7}

\bibitem[{{de Wit} {et~al.}(2008){de Wit}, {Oudmaijer}, {Fujiyoshi}, {Hoare}, {Honda}, {Kataza}, {Miyata}, {Okamoto}, {Onaka}, {Sako}, \& {Yamashita}}]{deWitetal.2008}
{de Wit}, W.~J., {Oudmaijer}, R.~D., {Fujiyoshi}, T., {et~al.} 2008, Astrophysical Journal Letters, 685, L75, \dodoi{10.1086/592384}

\bibitem[{{Decin} {et~al.}(2012){Decin}, {Cox}, {Royer}, {Van Marle}, {Vandenbussche}, {Ladjal}, {Kerschbaum}, {Ottensamer}, {Barlow}, {Blommaert}, {Gomez}, {Groenewegen}, {Lim}, {Swinyard}, {Waelkens}, \& {Tielens}}]{Decinetal.2012}
{Decin}, L., {Cox}, N.~L.~J., {Royer}, P., {et~al.} 2012, Astronomy and Astrophysics, 548, A113, \dodoi{10.1051/0004-6361/201219792}

\bibitem[{{Dolan} {et~al.}(2016){Dolan}, {Mathews}, {Lam}, {Quynh Lan}, {Herczeg}, \& {Dearborn}}]{Dolanetal.2016}
{Dolan}, M.~M., {Mathews}, G.~J., {Lam}, D.~D., {et~al.} 2016, Astrophysical Journal, 819, 7, \dodoi{10.3847/0004-637X/819/1/7}

\bibitem[{{Dorch} \& {Freytag}(2003)}]{DorchFreytag.2003}
{Dorch}, S.~B.~F., \& {Freytag}, B. 2003, in Modelling of Stellar Atmospheres, ed. N.~{Piskunov}, W.~W. {Weiss}, \& D.~F. {Gray}, Vol. 210, A12.
\newblock \doarXiv{astro-ph/0208523}

\bibitem[{{Ducati}(2002)}]{CSPCS.2002}
{Ducati}, J.~R. 2002, {VizieR Online Data Catalog: Catalogue of Stellar Photometry in Johnson's 11-color system.}

\bibitem[{Dupree {et~al.}(1987)Dupree, Baliunas, Hartmann, Nassiopoulos, Guinan, \& Sonneborn}]{Dupreeetal.1987}
Dupree, A., Baliunas, S., Hartmann, L., {et~al.} 1987, Astrophysical Journal, 317, \dodoi{10.1086/184917}

\bibitem[{{Dupree}(2011)}]{Dupree.2011}
{Dupree}, A.~K. 2011, in Physics of Sun and Star Spots, ed. D.~{Prasad Choudhary} \& K.~G. {Strassmeier}, Vol. 273, 188--194, \dodoi{10.1017/S1743921311015225}

\bibitem[{{Dupree} {et~al.}(2020){Dupree}, {Strassmeier}, {Matthews}, {Uitenbroek}, {Calderwood}, {Granzer}, {Guinan}, {Leike}, {Montarg{\`e}s}, {Richards}, {Wasatonic}, \& {Weber}}]{Dupreeetal.2020}
{Dupree}, A.~K., {Strassmeier}, K.~G., {Matthews}, L.~D., {et~al.} 2020, Astrophysical Journal, 899, 68, \dodoi{10.3847/1538-4357/aba516}

\bibitem[{{Dyck} {et~al.}(1992){Dyck}, {Benson}, {Ridgway}, \& {Dixon}}]{Dycketal.1992}
{Dyck}, H.~M., {Benson}, J.~A., {Ridgway}, S.~T., \& {Dixon}, D.~J. 1992, Astronomical Journal, 104, 1982, \dodoi{10.1086/116373}

\bibitem[{{Elias} {et~al.}(1985){Elias}, {Frogel}, \& {Humphreys}}]{Eliasetal.1985}
{Elias}, J.~H., {Frogel}, J.~A., \& {Humphreys}, R.~M. 1985, Astrophysical Journal, 57, 91, \dodoi{10.1086/190997}

\bibitem[{{Erdim} \& {H{\"u}daverdi}(2019)}]{SOAD2019}
{Erdim}, M.~K., \& {H{\"u}daverdi}, M. 2019, in American Institute of Physics Conference Series, Vol. 2178, Turkish Physical Society 35th International Physics Congress (TPS35) (AIP), 030023, \dodoi{10.1063/1.5135421}

\bibitem[{{FitzGerald}(1970)}]{Fitzgerald.1970}
{FitzGerald}, M.~P. 1970, Astronomy and Astrophysics, 4, 234

\bibitem[{Fitzpatrick(1999)}]{Fitzpatrick.1999}
Fitzpatrick, E.~L. 1999, Publications of the Astronomical Society of the Pacific, 111, 63, \dodoi{10.1086/316293}

\bibitem[{{Fitzpatrick}(2004)}]{Fitzpatrick.2004}
{Fitzpatrick}, E.~L. 2004, in Astronomical Society of the Pacific Conference Series, Vol. 309, Astrophysics of Dust, ed. A.~N. {Witt}, G.~C. {Clayton}, \& B.~T. {Draine}, 33.
\newblock \doarXiv{astro-ph/0401344}

\bibitem[{{Gezari} {et~al.}(1999){Gezari}, {Pitts}, \& {Schmitz}}]{CIO.1999}
{Gezari}, D.~Y., {Pitts}, P.~S., \& {Schmitz}, M. 1999, {VizieR Online Data Catalog: Catalog of Infrared Observations, Edition 5 (Gezari+ 1999)}

\bibitem[{{Goldberg} {et~al.}(2020){Goldberg}, {Bildsten}, \& {Paxton}}]{Goldbergetal.2020}
{Goldberg}, J.~A., {Bildsten}, L., \& {Paxton}, B. 2020, Astrophysical Journal, 891, 15, \dodoi{10.3847/1538-4357/ab7205}

\bibitem[{Goldberg {et~al.}(2022)Goldberg, Jiang, \& Bildsten}]{Goldberg_2022}
Goldberg, J.~A., Jiang, Y.-F., \& Bildsten, L. 2022, The Astrophysical Journal, 929, 156, \dodoi{10.3847/1538-4357/ac5ab3}

\bibitem[{Goldberg {et~al.}(2024)Goldberg, Joyce, \& Molnár}]{Goldberg_2024}
Goldberg, J.~A., Joyce, M., \& Molnár, L. 2024, The Astrophysical Journal, 977, 35, \dodoi{10.3847/1538-4357/ad87f4}

\bibitem[{Goldberg(1984)}]{Goldberg.1984}
Goldberg, L. 1984, Publications of the Astronomical Society of the Pacific, 96, 366, \dodoi{10.1086/131347}

\bibitem[{{Guinan} {et~al.}(2020){Guinan}, {Wasatonic}, {Calderwood}, \& {Carona}}]{Guinan.2020}
{Guinan}, E., {Wasatonic}, R., {Calderwood}, T., \& {Carona}, D. 2020, The Astronomer's Telegram, 13410-13512, 3

\bibitem[{{Guinan}(1984)}]{Guin.1984}
{Guinan}, E.~F. 1984, {Multiband Photoelectric Photometry of Betelgeuse}, Vol. 193 (Springer), 336--341, \dodoi{10.1007/3-540-12907-3_225}

\bibitem[{{Guinan} {et~al.}(2019){Guinan}, {Wasatonic}, \& {Calderwood}}]{Guinan.2019}
{Guinan}, E.~F., {Wasatonic}, R.~J., \& {Calderwood}, T.~J. 2019, The Astronomer's Telegram, 13341-13365, 2

\bibitem[{{Guo} \& {Li}(2002)}]{GuoLi.2002}
{Guo}, J.~H., \& {Li}, Y. 2002, Astrophysical Journal, 565, 559, \dodoi{10.1086/324295}

\bibitem[{{Harper} {et~al.}(2008){Harper}, {Brown}, \& {Guinan}}]{Harperetal.2008}
{Harper}, G.~M., {Brown}, A., \& {Guinan}, E.~F. 2008, Astronomical Journal, 135, 1430, \dodoi{10.1088/0004-6256/135/4/1430}

\bibitem[{{Harper} {et~al.}(2017){Harper}, {Brown}, {Guinan}, {O'Gorman}, {Richards}, {Kervella}, \& {Decin}}]{Harperetal.2017}
{Harper}, G.~M., {Brown}, A., {Guinan}, E.~F., {et~al.} 2017, Astronomical Journal, 154, 11, \dodoi{10.3847/1538-3881/aa6ff9}

\bibitem[{Harper {et~al.}(2020)Harper, Guinan, Wasatonic, \& Ryde}]{Harper_2020}
Harper, G.~M., Guinan, E.~F., Wasatonic, R., \& Ryde, N. 2020, The Astrophysical Journal, 905, 34, \dodoi{10.3847/1538-4357/abc1f0}

\bibitem[{Harris {et~al.}(2020)Harris, Millman, Van Der~Walt, Gommers, Virtanen, Cournapeau, Wieser, Taylor, Berg, Smith, {et~al.}}]{Numpy2020}
Harris, C.~R., Millman, K.~J., Van Der~Walt, S.~J., {et~al.} 2020, Nature, 585, 357

\bibitem[{{Haubois} {et~al.}(2009){Haubois}, {Perrin}, {Lacour}, {Verhoelst}, {Meimon}, {Mugnier}, {Thi{\'e}baut}, {Berger}, {Ridgway}, {Monnier}, {Millan-Gabet}, \& {Traub}}]{Hauboisetal.2009}
{Haubois}, X., {Perrin}, G., {Lacour}, S., {et~al.} 2009, Astronomy and Astrophysics, 508, 923, \dodoi{10.1051/0004-6361/200912927}

\bibitem[{Hick {et~al.}(2007)Hick, Buffington, \& Jackson}]{Hicketal.2007}
Hick, P., Buffington, A., \& Jackson, B.~V. 2007, in Solar Physics and Space Weather Instrumentation II, ed. S.~Fineschi \& R.~A. Viereck, Vol. 6689, International Society for Optics and Photonics (SPIE), 85 -- 92, \dodoi{10.1117/12.734808}

\bibitem[{{Hindsley} \& {Harrington}(1994)}]{HindHarr.1994}
{Hindsley}, R.~B., \& {Harrington}, R.~S. 1994, {The U.S. Naval Observatory Catalog of Positions of Infrared Stellar Sources}, \dodoi{10.1086/116852}

\bibitem[{{Hoffleit} \& {Warren}(1995)}]{HoffWarr.1991}
{Hoffleit}, D., \& {Warren}, W.~H., J. 1995, VizieR Online Data Catalog, V/50

\bibitem[{{Hunter}(2007)}]{Matplotlib2007}
{Hunter}, J.~D. 2007, Computing in Science and Engineering, 9, 90, \dodoi{10.1109/MCSE.2007.55}

\bibitem[{{Jackson} {et~al.}(2004){Jackson}, {Buffington}, {Hick}, {Altrock}, {Figueroa}, {Holladay}, {Johnston}, {Kahler}, {Mozer}, {Price}, {Radick}, {Sagalyn}, {Sinclair}, {Simnett}, {Eyles}, {Cooke}, {Tappin}, {Kuchar}, {Mizuno}, {Webb}, {Anderson}, {Keil}, {Gold}, \& {Waltham}}]{Jackson.2004}
{Jackson}, B.~V., {Buffington}, A., {Hick}, P.~P., {et~al.} 2004, {The Solar Mass-Ejection Imager (SMEI) Mission}, \dodoi{10.1007/s11207-004-2766-3}

\bibitem[{{Jayasinghe} {et~al.}(2018){Jayasinghe}, {Stanek}, {Kochanek}, {Shappee}, {Holoien}, {Thompson}, {Prieto}, {Dong}, {Pawlak}, {Pejcha}, {Shields}, {Pojmanski}, {Otero}, {Britt}, \& {Will}}]{ASASSN.2019}
{Jayasinghe}, T., {Stanek}, K.~Z., {Kochanek}, C.~S., {et~al.} 2018, {The ASAS-SN catalogue of variable stars - II. Uniform classification of 412 000 known variables}, \dodoi{10.1093/mnras/stz844}

\bibitem[{{Josselin} \& {Plez}(2007)}]{JosselinPlez.2007}
{Josselin}, E., \& {Plez}, B. 2007, Astronomy and Astrophysics, 469, 671, \dodoi{10.1051/0004-6361:20066353}

\bibitem[{{Joyce} {et~al.}(2020){Joyce}, {Leung}, {Moln{\'a}r}, {Ireland}, {Kobayashi}, \& {Nomoto}}]{Joyceetal.2020}
{Joyce}, M., {Leung}, S.-C., {Moln{\'a}r}, L., {et~al.} 2020, Astrophysical Journal, 902, 63, \dodoi{10.3847/1538-4357/abb8db}

\bibitem[{Kafka(2020)}]{Kafka.2020}
Kafka, S. 2020, {AAVSO International Database}.
\newblock \url{https://www.aavso.org/}

\bibitem[{{Karovska} {et~al.}(1986){Karovska}, {Nisenson}, \& {Noyes}}]{Karovska.1986}
{Karovska}, M., {Nisenson}, P., \& {Noyes}, R. 1986, \apj, 308, 260, \dodoi{10.1086/164497}

\bibitem[{{Kiss} {et~al.}(2006){Kiss}, {Szab{\'o}}, \& {Bedding}}]{Kissetal.2006}
{Kiss}, L.~L., {Szab{\'o}}, G.~M., \& {Bedding}, T.~R. 2006, Monthly Notices of the Royal Astronomical Society, 372, 1721, \dodoi{10.1111/j.1365-2966.2006.10973.x}

\bibitem[{{Kochanek} {et~al.}(2012){Kochanek}, {Khan}, \& {Dai}}]{Kochaneketal.2012}
{Kochanek}, C.~S., {Khan}, R., \& {Dai}, X. 2012, Astrophysical Journal, 759, 20, \dodoi{10.1088/0004-637X/759/1/20}

\bibitem[{{Lambert} {et~al.}(1984){Lambert}, {Brown}, {Hinkle}, \& {Johnson}}]{Lambertetal.1984}
{Lambert}, D.~L., {Brown}, J.~A., {Hinkle}, K.~H., \& {Johnson}, H.~R. 1984, Astrophysical Journal, 284, 223, \dodoi{10.1086/162401}

\bibitem[{Leavitt(1912)}]{Leavitt.1912}
Leavitt, H.~S. 1912, Harvard Circ., 173, 1 (rep. by E. C. Pickering)

\bibitem[{{Lenz} \& {Breger}(2005)}]{LenzBreger.2005}
{Lenz}, P., \& {Breger}, M. 2005, Communications in Asteroseismology, 146, 53, \dodoi{10.1553/cia146s53}

\bibitem[{{Levesque}(2018)}]{Levesque.2018}
{Levesque}, E.~M. 2018, Astrophysical Journal, 867, 155, \dodoi{10.3847/1538-4357/aae776}

\bibitem[{{Levesque} \& {Massey}(2020)}]{LevesqueMessey.2020}
{Levesque}, E.~M., \& {Massey}, P. 2020, Astrophysical Journal Letters, 891, L37, \dodoi{10.3847/2041-8213/ab7935}

\bibitem[{{Lomb}(1976)}]{Lomb.1976}
{Lomb}, N.~R. 1976, Astrophysics and Space Sciences, 39, 447, \dodoi{10.1007/BF00648343}

\bibitem[{{L{\'o}pez Ariste} {et~al.}(2018){L{\'o}pez Ariste}, {Mathias}, {Tessore}, {L{\`e}bre}, {Auri{\`e}re}, {Petit}, {Ikhenache}, {Josselin}, {Morin}, \& {Montarg{\`e}s}}]{LopezAristeetal.2018}
{L{\'o}pez Ariste}, A., {Mathias}, P., {Tessore}, B., {et~al.} 2018, Astronomy and Astrophysics, 620, A199, \dodoi{10.1051/0004-6361/201834178}

\bibitem[{{Lovy} {et~al.}(1984){Lovy}, {Maeder}, {Noels}, \& {Gabriel}}]{Lovyetal.1984}
{Lovy}, D., {Maeder}, A., {Noels}, A., \& {Gabriel}, M. 1984, Astronomy and Astrophysics Journal, 133, 307

\bibitem[{{Mackey} {et~al.}(2013){Mackey}, {Mohamed}, {Neilson}, {Langer}, \& {Meyer}}]{Mackeyetal.2013}
{Mackey}, J., {Mohamed}, S., {Neilson}, H.~R., {Langer}, N., \& {Meyer}, D.~M.~A. 2013, in EAS Publications Series, Vol.~60, EAS Publications Series, ed. P.~{Kervella}, T.~{Le Bertre}, \& G.~{Perrin}, 253--259, \dodoi{10.1051/eas/1360029}

\bibitem[{MacLeod {et~al.}(2024)MacLeod, Blunt, Rosa, Dupree, Granzer, Harper, Huang, Leiner, Loeb, Nielsen, Strassmeier, Wang, \& Weber}]{MacLeod_2025}
MacLeod, M., Blunt, S., Rosa, R. J.~D., {et~al.} 2024, The Astrophysical Journal, 978, 50, \dodoi{10.3847/1538-4357/ad93c8}

\bibitem[{{Massey} \& {Evans}(2016)}]{MasseyEvans.2016}
{Massey}, P., \& {Evans}, K.~A. 2016, Astrophysical Journal, 826, 224, \dodoi{10.3847/0004-637X/826/2/224}

\bibitem[{Massey {et~al.}(2007)Massey, Levesque, Plez, \& Olsen}]{Masseyetal.2007}
Massey, P., Levesque, E.~M., Plez, B., \& Olsen, K. A.~G. 2007, Proceedings of the International Astronomical Union, 3, 97–110, \dodoi{10.1017/s1743921308020383}

\bibitem[{{Massey} {et~al.}(2005){Massey}, {Plez}, {Levesque}, {Olsen}, {Clayton}, \& {Josselin}}]{Masseyetal.2005}
{Massey}, P., {Plez}, B., {Levesque}, E.~M., {et~al.} 2005, Astrophysical Journal, 634, 1286, \dodoi{10.1086/497065}

\bibitem[{{Massey} {et~al.}(2009){Massey}, {Silva}, {Levesque}, {Plez}, {Olsen}, {Clayton}, {Meynet}, \& {Maeder}}]{Masseyetal.2009}
{Massey}, P., {Silva}, D.~R., {Levesque}, E.~M., {et~al.} 2009, Astrophysical Journal, 703, 420, \dodoi{10.1088/0004-637X/703/1/420}

\bibitem[{{Mattila} {et~al.}(2012){Mattila}, {Dahlen}, {Efstathiou}, {Kankare}, {Melinder}, {Alonso-Herrero}, {P{\'e}rez-Torres}, {Ryder}, {V{\"a}is{\"a}nen}, \& {{\"O}stlin}}]{Mattilaetal.2012}
{Mattila}, S., {Dahlen}, T., {Efstathiou}, A., {et~al.} 2012, Astrophysical Journal, 756, 111, \dodoi{10.1088/0004-637X/756/2/111}

\bibitem[{{Mauron} \& {Josselin}(2011)}]{MauronJosselin.2011}
{Mauron}, N., \& {Josselin}, E. 2011, Astronomy and Astrophysics, 526, A156, \dodoi{10.1051/0004-6361/201013993}

\bibitem[{McKinney {et~al.}(2011)}]{Pandas2011}
McKinney, W., {et~al.} 2011, Python for high performance and scientific computing, 14, 1

\bibitem[{{Mermilliod}(1987)}]{Mermilliod.1987}
{Mermilliod}, J.~C. 1987, {UBV Photoelectric Photometry Catalogue (1986): I. The Original data}

\bibitem[{{Mermilliod}(2006)}]{Mermilliod.2006}
---. 2006, {Homogeneous Means in the UBV System}

\bibitem[{{Meynet} {et~al.}(2015){Meynet}, {Chomienne}, {Ekstr{\"o}m}, {Georgy}, {Granada}, {Groh}, {Maeder}, {Eggenberger}, {Levesque}, \& {Massey}}]{Meynetetal.2015}
{Meynet}, G., {Chomienne}, V., {Ekstr{\"o}m}, S., {et~al.} 2015, Astronomy and Astrophysics, 575, A60, \dodoi{10.1051/0004-6361/201424671}

\bibitem[{{Mohamed} {et~al.}(2012){Mohamed}, {Mackey}, \& {Langer}}]{Mohamedetal.2012}
{Mohamed}, S., {Mackey}, J., \& {Langer}, N. 2012, Astronomy and Astrophysics, 541, A1, \dodoi{10.1051/0004-6361/201118002}

\bibitem[{{Nakaya} {et~al.}(2001){Nakaya}, {Watanabe}, {Ando}, {Nagata}, \& {Sato}}]{Nakayaetal.2001}
{Nakaya}, H., {Watanabe}, M., {Ando}, M., {Nagata}, T., \& {Sato}, S. 2001, Astronomical Journal, 122, 876, \dodoi{10.1086/321178}

\bibitem[{{Neugent} {et~al.}(2020){Neugent}, {Massey}, {Georgy}, {Drout}, {Mommert}, {Levesque}, {Meynet}, \& {Ekstr{\"o}m}}]{Neugentetal.2020}
{Neugent}, K.~F., {Massey}, P., {Georgy}, C., {et~al.} 2020, Astrophysical Journal, 889, 44, \dodoi{10.3847/1538-4357/ab5ba0}

\bibitem[{{Nicolet}(1978)}]{Nicolet.1978}
{Nicolet}, B. 1978, Astronomy and Astrophysicss, 34, 1

\bibitem[{{Ohnaka} {et~al.}(2009){Ohnaka}, {Hofmann}, {Benisty}, {Chelli}, {Driebe}, {Millour}, {Petrov}, {Schertl}, {Stee}, {Vakili}, \& {Weigelt}}]{Ohnakaetal.2009}
{Ohnaka}, K., {Hofmann}, K.~H., {Benisty}, M., {et~al.} 2009, Astronomy and Astrophysics, 503, 183, \dodoi{10.1051/0004-6361/200912247}

\bibitem[{{Ohnaka} {et~al.}(2011){Ohnaka}, {Weigelt}, {Millour}, {Hofmann}, {Driebe}, {Schertl}, {Chelli}, {Massi}, {Petrov}, \& {Stee}}]{Ohnakaetal.2011}
{Ohnaka}, K., {Weigelt}, G., {Millour}, F., {et~al.} 2011, Astronomy and Astrophysics, 529, A163, \dodoi{10.1051/0004-6361/201016279}

\bibitem[{{Percy} \& {Deibert}(2016)}]{PercyDeibert.2016}
{Percy}, J.~R., \& {Deibert}, E. 2016, Journal of the American Association of Variable Star Observers (JAAVSO), 44, 94.
\newblock \doarXiv{1607.06482}

\bibitem[{{Perrin} {et~al.}(2004){Perrin}, {Ridgway}, {Coud{\'e} du Foresto}, {Mennesson}, {Traub}, \& {Lacasse}}]{Perrinetal.2004}
{Perrin}, G., {Ridgway}, S.~T., {Coud{\'e} du Foresto}, V., {et~al.} 2004, Astronomy and Astrophysics, 418, 675, \dodoi{10.1051/0004-6361:20040052}

\bibitem[{{Price} {et~al.}(2010){Price}, {Smith}, {Kuchar}, {Mizuno}, \& {Kraemer}}]{NIRLC.2010}
{Price}, S.~D., {Smith}, B.~J., {Kuchar}, T.~A., {Mizuno}, D.~R., \& {Kraemer}, K.~E. 2010, {3.6 Years of DIRBE Near-infrared Stellar Light Curves}, \dodoi{10.1088/0067-0049/190/2/203}

\bibitem[{{Scargle}(1982)}]{Scargle.1982}
{Scargle}, J.~D. 1982, Astrophysical Journal, 263, 835, \dodoi{10.1086/160554}

\bibitem[{{Schlegel} {et~al.}(1998){Schlegel}, {Finkbeiner}, \& {Davis}}]{Schlegeletal.1998}
{Schlegel}, D.~J., {Finkbeiner}, D.~P., \& {Davis}, M. 1998, Astrophysical Journal, 500, 525, \dodoi{10.1086/305772}

\bibitem[{{Schwarzschild}(1975)}]{Schwarzschild.1975}
{Schwarzschild}, M. 1975, Astrophysical Journal, 195, 137, \dodoi{10.1086/153313}

\bibitem[{{Shappee} {et~al.}(2014){Shappee}, {Prieto}, {Grupe}, {Kochanek}, {Stanek}, {De Rosa}, {Mathur}, {Zu}, {Peterson}, {Pogge}, {Komossa}, {Im}, {Jencson}, {Holoien}, {Basu}, {Beacom}, {Szczygie{\l}}, {Brimacombe}, {Adams}, {Campillay}, {Choi}, {Contreras}, {Dietrich}, {Dubberley}, {Elphick}, {Foale}, {Giustini}, {Gonzalez}, {Hawkins}, {Howell}, {Hsiao}, {Koss}, {Leighly}, {Morrell}, {Mudd}, {Mullins}, {Nugent}, {Parrent}, {Phillips}, {Pojmanski}, {Rosing}, {Ross}, {Sand}, {Terndrup}, {Valenti}, {Walker}, \& {Yoon}}]{ASASSN.2014}
{Shappee}, B.~J., {Prieto}, J.~L., {Grupe}, D., {et~al.} 2014, {The Man behind the Curtain: X-Rays Drive the UV through NIR Variability in the 2013 Active Galactic Nucleus Outburst in NGC 2617}, \dodoi{10.1088/0004-637X/788/1/48}

\bibitem[{{Smith} {et~al.}(2004){Smith}, {Price}, \& {Baker}}]{DIBREPSC.2004}
{Smith}, B.~J., {Price}, S.~D., \& {Baker}, R.~I. 2004, {The COBE DIRBE Point Source Catalog}, \dodoi{10.1086/423248}

\bibitem[{{Soszy{\'n}ski} \& {Udalski}(2014)}]{SoszynskiUdalski.2014}
{Soszy{\'n}ski}, I., \& {Udalski}, A. 2014, Astrophysical Journal, 788, 13, \dodoi{10.1088/0004-637X/788/1/13}

\bibitem[{{Stothers} \& {Leung}(1971)}]{StothersLeung.1971}
{Stothers}, R., \& {Leung}, K.~C. 1971, Astronomy and Astrophysics, 10, 290

\bibitem[{{Stothers}(2010)}]{Stothers.2010}
{Stothers}, R.~B. 2010, Astrophysical Journal, 725, 1170, \dodoi{10.1088/0004-637X/725/1/1170}

\bibitem[{{Tabur} {et~al.}(2009){Tabur}, {Kiss}, \& {Bedding}}]{Taburetal.2009}
{Tabur}, V., {Kiss}, L.~L., \& {Bedding}, T.~R. 2009, Astrophysical Journal Letters, 703, L72, \dodoi{10.1088/0004-637X/703/1/L72}

\bibitem[{{van Leeuwen}(2007)}]{vanLeeuwen.2007}
{van Leeuwen}, F. 2007, Astronomy and Astrophysics, 474, 653, \dodoi{10.1051/0004-6361:20078357}

\bibitem[{{Verhoelst} {et~al.}(2009){Verhoelst}, {van der Zypen}, {Hony}, {Decin}, {Cami}, \& {Eriksson}}]{Verhoelstetal.2009}
{Verhoelst}, T., {van der Zypen}, N., {Hony}, S., {et~al.} 2009, Astronomy and Astrophysics, 498, 127, \dodoi{10.1051/0004-6361/20079063}

\bibitem[{Virtanen {et~al.}(2020)Virtanen, Gommers, Oliphant, Haberland, Reddy, Cournapeau, Burovski, Peterson, Weckesser, Bright, {van der Walt}, Brett, Wilson, Millman, Mayorov, Nelson, Jones, Kern, Larson, Carey, Polat, Feng, Moore, {VanderPlas}, Laxalde, Perktold, Cimrman, Henriksen, Quintero, Harris, Archibald, Ribeiro, Pedregosa, {van Mulbregt}, \& {SciPy 1.0 Contributors}}]{Scipy2020}
Virtanen, P., Gommers, R., Oliphant, T.~E., {et~al.} 2020, Nature Methods, 17, 261, \dodoi{10.1038/s41592-019-0686-2}

\bibitem[{{Webb} {et~al.}(2006){Webb}, {Mizuno}, {Buffington}, {Cooke}, {Eyles}, {Fry}, {Gentile}, {Hick}, {Holladay}, {Howard}, {Hewitt}, {Jackson}, {Johnston}, {Kuchar}, {Mozer}, {Price}, {Radick}, {Simnett}, \& {Tappin}}]{Webbetal.2006}
{Webb}, D.~F., {Mizuno}, D.~R., {Buffington}, A., {et~al.} 2006, Journal of Geophysical Research (Space Physics), 111, A12101, \dodoi{10.1029/2006JA011655}

\bibitem[{{Ye{\c{s}}ilyaprak} \& {Aslan}(2004)}]{Aslan.2004}
{Ye{\c{s}}ilyaprak}, C., \& {Aslan}, Z. 2004, Monthly Notices of the Royal Astronomical Society, 355, 601, \dodoi{10.1111/j.1365-2966.2004.08344.x}

\bibitem[{{Yoon} \& {Cantiello}(2010)}]{YoonCantiello.2010}
{Yoon}, S.-C., \& {Cantiello}, M. 2010, Astrophysical Journal Letters, 717, L62, \dodoi{10.1088/2041-8205/717/1/L62}

\end{thebibliography}
\bibliographystyle{aasjournal}

\end{document}